\documentclass[runningheads]{llncs}
\usepackage[T1]{fontenc}
\usepackage{graphicx}
\usepackage{array}
\usepackage{tabularx}
\usepackage{adjustbox}
\usepackage[table]{xcolor}
\usepackage{tcolorbox} 
\usepackage{cite}
\usepackage{hyperref}
\usepackage{todonotes}
\usepackage{listings}
\usepackage{amsmath}
\usepackage{marginnote}

\newif\ifrev
\revfalse
\ifrev
\newcommand{\R}[2]{%
  \marginnote{\textcolor{gray}{#1}}%
  \textcolor{blue}{#2}%
 }
\else
  \newcommand{\R}[2]{#2}
\fi

\begin{document}
%
\title{Fairness in Healthcare Processes: A Quantitative Analysis of Decision Making in Triage}
\titlerunning{Fairness in Healthcare Processes}
%
\author{Rachmadita Andreswari\inst{1,2,4},
Stephan A. Fahrenkrog-Petersen\inst{2,3}, \\
Jan Mendling\inst{1,2}}
\authorrunning{R. Andreswari et al.}
%
\institute{Humboldt-Universität zu Berlin, Rudower Chaussee 25, 12489 Berlin, Germany \and
Weizenbaum Institut, Hardenbergstraße 32, 10623 Berlin, Germany \and
University of Liechtenstein, Fürst-Franz-Josef-Strasse,
9490 Vaduz, Liechtenstein \and
Telkom University, Jl Telekomunikasi 1, 40257 Bandung, Indonesia
\email{\{rachmadita.andre.swari,jan.mendling\}@hu-berlin.de, stephan.fahrenkrog@uni.li}}
\maketitle              
%
\begin{abstract}


Fairness in automated decision-making has become a critical concern, particularly in high-pressure healthcare scenarios such as emergency triage, where fast and equitable decisions are essential. Process mining is increasingly investigating fairness. There is a growing area focusing on  fairness-aware algorithms. So far, we know less how these concepts perform on  empirical healthcare data or how they cover aspects of justice theory. This study addresses this research problem and proposes a process mining approach to assess fairness in triage by linking real-life event logs with conceptual dimensions of justice. Using the MIMICEL event log 
we analyze time, re-do, deviation and decision as process outcomes, and evaluate the influence of age, gender, race, language and insurance using the Kruskal–Wallis, Chi-square and effect size measurements. These outcomes are mapped to justice dimensions to support the development of a conceptual framework. The results demonstrate which aspects of potential unfairness in high-acuity and sub-acute surface. In this way, this study contributes empirical insights that support further research in responsible, fairness-aware process mining in healthcare.

\end{abstract}

\section{Introduction}
\label{sec:introduction}
Healthcare systems often involve complex business processes due to the variety of cases and diseases they address. These processes have been widely explored using process mining as a key technique to gain insights into what occurs within healthcare workflows \cite{rojas2016process}. Several studies on process mining in healthcare have addressed problems related to medical services such as emergency care, intensive care units, medication management, and the treatment of various diseases \cite{rojas2016process}. 
\R{2.1}{These healthcare processes are instances of process-aware information systems (PAIS) \cite{van2009process}, where event logs enable empirical analysis of operational workflows. The relevance of fairness as an IS research topic has been recognized since early work on equity perceptions in MIS contexts \cite{joshi1989measurement}, and this study extends that tradition to process-level analysis of healthcare workflows.}

Ethical concerns are a critical consideration of medical decision-making. One of the most important factors here is fairness, which means ensuring unbiased treatment for all patients \cite{cappelen2006responsibility}.
In this context, fairness means that decisions are based on objective factors such as vital signs \cite{sapra2023vital} and medical needs\cite{varkey2021principles}. Nonetheless, subjective factors such as race \cite{chou2007gender}, age \cite{buja2014need}, social status \cite{chakraborty2024setting}, and gender \cite{chou2007gender} may influence clinical decisions. This influence may lead to a loss of treatment priority for certain patients, potentially resulting in severe health consequences or even the loss of hope for recovery.

In recent years, process mining is increasingly investigating fairness. There is a growing area focusing on fairness-aware algorithms~\cite{peeperkorn2025achieving}, \cite{muskan2024extending}, \cite{de2024achieving}, \R{3.1}{\cite{dorleon2023fapfid}} based on conceptual foundations of, e.g., \cite{pohl2022discrimination,Mannhardt2022Responsible}. So far, we know less how these concepts perform on empirical healthcare data or how they cover aspects of justice theory. \R{3.2}{In medical systems, sensitive attributes such as age, gender, race, insurance status, and language are legally protected under non-discrimination frameworks such as Emergency Medical Treatment and Active Labor Act (EMTALA)\cite{zibulewsky2001emergency}, yet some may legitimately influence clinical decisions based on established medical standards, such as age-related health conditions. This duality makes fairness assessment in healthcare particularly challenging, as distinguishing clinically justified from socially unjustified disparities requires careful empirical analysis.} 
This study addresses this research problem. We focus on triage processes because unfair handling has been reported in recent studies \cite{lauridsen2020emergency}. We propose a process mining approach to assess fairness using the public event data from the MIMIC-IV Emergency Department database \cite{johnson2023mimic} in three steps of (i) operationalizing triage decisions in relation to business process outcomes; (ii) mapping process outcomes to MIMIC event logs; and (iii) analyzing fairness using a statistical analysis utilizing the Kruskal-Wallis, Chi-square, epsilon-squared, and Cramér’s V tests. 
In this way, we obtain insights into the significance of sensitive attributes for determining outcomes.

The paper is structured as follows: Section~\ref{sec:background} discusses the background of this study, including recent research on process mining in healthcare and the role of fairness in business process management. Section~\ref{sec:method} describes our research methodology for deriving insights from empirical MIMIC-IV ED data. Section~\ref{sec:result} presents the results and a discussion on implications for healthcare process mining. Section~\ref{sec:conclusion} summarizes our findings and outlines directions for future research.

\vspace{-1em}
\section{Background}
\label{sec:background}
\vspace{-1em}
This section reviews related research on triage, fairness, business process management and process mining. It examines the general triage process and its challenges, the importance of fairness in triage decision-making, how BPM and process mining has been previously implemented to address these issues.

\subsection{Triage and Challenges}
\begin{figure}
    \centering
    \resizebox{\textwidth}{!}{\includegraphics{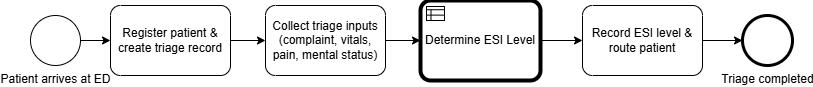}}
    \caption{ESI Triage Process as BPMN\cite{gilboy2012emergency}}
    \label{fig:ESI}
\end{figure}

Triage is a fundamental process in emergency departments and is implemented across healthcare units worldwide. Several methods are used in triage, including the Emergency Severity Index (ESI), the Manchester Triage System (MTS), 
and, for mass-casualty incidents, Simple Triage and Rapid Treatment (START) and variants \cite{iserson2007triage}. 
The triage method adopted by Beth Israel Deaconess Medical Center (BIDMC) in the United States \cite{fernandes2020predicting} uses ESI. This method classifies patients based on the urgency of their condition. Figure~\ref{fig:ESI} shows the ESI process and Table \ref{tab:dmn_initial_esi} how patients are assessed and categorized into five levels (1–5). 

\vspace{-2em}

\begin{table}[htbp]
\centering
\caption{DMN Decision Table for Initial ESI Level \cite{gilboy2012emergency}}
\label{tab:dmn_initial_esi}
\resizebox{0.7\textwidth}{!}{
\begin{tabular}{p{2.5cm} p{6cm} p{2.6cm} c}
\hline
\textbf{Life-saving} &
\textbf{High risk / confusion / severe pain} &
\textbf{Resources} &
\textbf{ESI} \\
\hline
Yes & -- & -- & 1 \\
No  & Yes & -- & 2 \\
No  & No  & Many & 3 \\
No  & No  & One  & 4 \\
No  & No  & None & 5 \\
\hline
\end{tabular}
}
\end{table}


As shown in Table~\ref{tab:dmn_initial_esi}, ESI level 1 requires immediate life-saving intervention, while levels 2--5 are determined by risk assessment, expected resource use, and vital signs \cite{gilboy2012emergency}.


\subsection{Studies on Fairness in Triage}
Although the ESI method provides a systematic framework, its implementation in emergency departments faces practical and ethical challenges.
Proper triage requires experienced healthcare professionals \cite{ahmed2024ethical}, as expertise and exposure to high-pressure situations play a critical role in making accurate and ethical decisions \cite{ghanbari2021fair}. Establishing clear, standardized prioritization criteria based on objective health assessments \cite{lauridsen2020emergency}, \cite{ahmed2024ethical} is crucial. If patient prioritization is influenced by social, cultural, or economic factors, it can lead to healthcare disparities, favoring privileged patients over others. Therefore, humanitarian values \cite{ahmed2024ethical} and the goal of saving the maximum number of lives \cite{ahmed2024ethical}, \cite{holzer2025role} must take precedence in triage assessment frameworks.

The various challenges can cause problems with fairness.
Judgments that rely on experience and intuition of medical staff can introduce cognitive biases and unintentional discrimination, especially when clinical information is incomplete 
In addition, the completeness and consistency of triage data in electronic medical records are often limited, which complicates data-driven process analysis and detection of inequities in care flow \cite{mashoufi2023data}. Time pressures and resource limitation in the emergency room can force rushed triage decisions and potentially affected the accuracy of patient prioritization \cite{soola2022evaluation}. Studies have shown that subjective factors such as social status \cite{chakraborty2024setting}, economic background \cite{holzer2025role}, age \cite{ahmed2024ethical}, and gender \cite{ahmed2024ethical}, \cite{holzer2025role} may also influence triage decisions. Each triage assessment recorded in the MIMIC-IV Emergency Department event log represents a process instance of this decision-making procedure, which forms the basis of our empirical analysis. These issues highlight why fairness in triage processes is a critical topic for empirical investigation.

Several measures have been discussed to address challenges.
Studies \cite{lauridsen2020emergency} indicate that prioritizing critical patients is relatively straightforward, whereas prioritizing non-critical cases is more complex. While a first-come, first-served queue system can be used \cite{lauridsen2020emergency}, \cite{ghanbari2021fair}, it does not always align with real-world emergencies, particularly when patients have different levels of urgency. In such cases, a lottery-based system \cite{ghanbari2021fair,broome1990fairness} may offer a fair alternative, ensuring equal treatment for patients with similar conditions while minimizing subjective biases.

Fair and equal treatment is \R{1.6}{regulated} for emergency care in the U.S. health system is regulated under EMTALA, which mandates nondiscriminatory triage and treatment irrespective of race, gender, insurance status, socioeconomic background, or language \cite{zibulewsky2001emergency}. The Emergency Severity Index (ESI) is similarly designed to prioritize clinical urgency over non-clinical characteristics. Therefore, deviations in waiting time or care trajectory across sensitive attributes should only reflect differences in clinical need. Unexplained deviations may indicate a breach in fairness principle.

\subsection{Fairness-Aware Process Mining}
In the context of process mining, research on responsible practices has been conducted by Mannhardt et al. \cite{Mannhardt2022Responsible}. They emphasized that process mining without responsibility may have a negative impact, particularly in cases of misuse of confidential data. They advocate to consider fairness, accuracy, confidentiality, and transparency (FACT) \cite{Aalst2017Responsible} when conducting responsible process mining research. Without fairness mechanisms~\cite{pohl2022discrimination}
, process analysis risks producing biased recommendations and violating fairness principles. 

Recent developments in fairness-aware process mining propose embedding discrimination metrics into process model to avoid unfair root-cause analyses. For instance, \cite{qafari2019fairness} suggests constructing classifiers that remove unwanted associations between sensitive attributes and outcomes to uncover implicit but practically relevant patterns. \R{1.5, 3.1}{More recent approaches extend this direction by integrating fairness constraints into predictive process monitoring \cite{peeperkorn2025achieving}, genetic process discovery \cite{muskan2024extending}, adversarial learning \cite{de2024achieving}, and and fairness-aware machine learning under imbalanced data conditions \cite{dorleon2023fapfid}. These approaches share a focus on algorithmic fairness, ensuring that predictive models or discovered process models do not discriminate across demographic groups.}

Ethical values such as fairness, accountability, transparency, and responsibility can be embedded throughout the BPM lifecycle to align process design and execution with ethical concern \cite{kern2024navigating}. Integrating these values not only enhances the validity of fairness assessments but also supports more ethical and socially legitimate business process implementations.

\R{1.1, 1.5}{In contrast, this paper takes a process-level fairness perspective: rather than building or evaluating predictive models, we diagnose disparities in how the triage process is actually executed across demographic groups using empirical event log data. While algorithmic fairness asks whether a model treats groups equitably, process-level fairness asks whether the process itself, in terms of time, re-do, deviation, and decision, operates equitably in practice.}

Fairness-aware process mining is an appropriate technique because unfairness can emerge from process structures and execution patterns, rather than solely from individual decisions. In triage, this makes process-level fairness analysis essential for exposing disparities that are invisible at the outcome level.

\section{Method}
\label{sec:method}

This section outlines the structured stages through which we move from empirical analysis to the conceptualization of fairness-aware process mining in triage settings. An overview of the data preparation and analysis pipeline is illustrated in Figure \ref{fig:processing}. We adopt a case study research approach \cite{runeson2009guidelines}. 
\vspace{-1em}
\subsection{Research Objective and Approach}


Our goal is to systematically extract fairness-relevant elements from healthcare event log data and organize them into fairness-oriented process mining dimensions. To that end, we build on Gilliand's model of organizational justice distinguishing distributive, procedural and interactional dimensions~\cite{gilliland1993perceived}.
In particular, we try to understand (a) how applicable PM techniques are for fairness assessment and (b) how plausible the results are. The case under investigation is the MIMIC-IV ED dataset on the triage process. Following guidelines from \cite{runeson2009guidelines}, we adopt an exploratory and explanatory approach to examine business process outcomes (BPOs) related to fairness. 
We combine process mining and non-parametric statistical tests to analyze outcomes across race, gender, age, insurance and language. The selected BPOs are then mapped to justice dimensions distributive, procedural and interactional justice as an analytical framework. 

\subsection{Data Collection and Processing}

\begin{figure}
    \centering
    \includegraphics[width=0.8\linewidth]{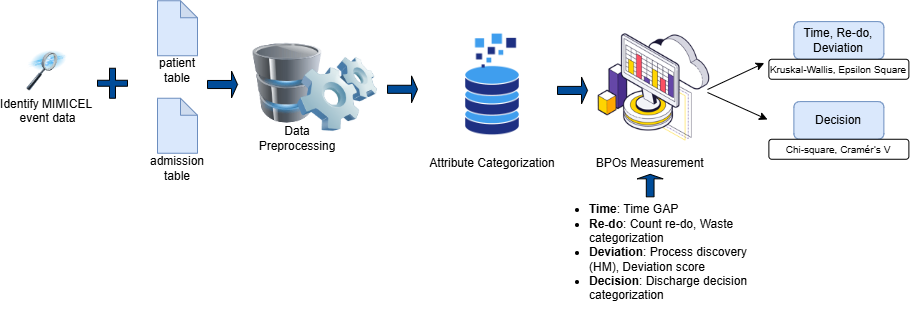}
    \caption{Data Processing Pipeline for Fairness Analysis in Triage}
    \label{fig:processing}
\end{figure}

MIMIC-IV ED is a large dataset comprising records of patients admitted to the emergency department or the intensive care unit at BIDMC, Boston, US \cite{johnson2023mimic}. It contains five primary tables: diagnosis, medrecon, pyxis, triage, and vitalsign. For this study, we utilize an event log conversion of the MIMIC-IV ED, known as MIMICEL \cite{wei2023mimicel}. MIMICEL includes essential columns required for process mining: case\_id, timestamp, and activity. It also provides case attributes represented by columns directly taken from the original MIMIC-IV ED tables. Figure \ref{fig:processing} shows our data processing pipeline to prepare the data for out study.

\vspace{-0.5em}

As we aim to analyze fairness based on patient severity, we first assess severity using the acuity level available in the MIMIC-IV ED dataset. This acuity assessment is derived from the ESI method \cite{fernandes2020predicting}. Although ESI is designed to guide prioritization objectively, it may be inconsistent in practice, particularly across demographic groups. We assume that treatment should be delivered consistently regardless of race, gender, age, insurance or language which forms the basis for examining potential disparities. 

The data were preprocessed by cleaning and imputing values for each variable within the same case ID, followed by validation after joining three tables: patients, admissions, and MIMICEL. The attributes were subsequently grouped, and consistency counts were computed for each category, resulting in a final table grouped by race, age, gender, insurance, and language. The categorization for each attribute is presented in \cite{andreswari2026fairnesshealthcareprocessesquantitative}. Each categorization is based on the closest corresponding group for each attribute. The “unknown” and “other” categories are retained, as some patients lack information on insurance, language, or race. For race, categories labeled as “deleted” are excluded.

Disparities were examined based on the selected business process outcomes (BPOs) for connections with demographic attributes. Statistical differences in outcomes across gender, age, race, insurance, and language were assessed using the Kruskal–Wallis and chi-square tests to compare multiple independent groups. \R{3.2, 3.5}{While several fairness metrics have been proposed for evaluating algorithmic fairness in predictive models \cite{garg2020fairness} these are designed to assess model outputs rather than process behavior. In this study, we adopt distributional statistical tests, namely the Kruskal-Wallis test and chi-square test, as they are more appropriate for comparing process outcome distributions across demographic groups without assuming a predictive model. Effect sizes ($\varepsilon^2$ and Cramér's V) are used as the primary measure of practical relevance, with statistical significance serving as a complementary indicator.}

Finally, we categorize these characteristics into justice dimensions such as distributive, procedural, and interactional justice as the basis for analyzing organizational justice in emergency healthcare processes. This conceptual structure supports our later analysis of fairness patterns and guides the integration of fairness into process mining techniques.

\vspace{-1em}
\subsection{Process Aspect Selection}
\vspace{-1em}

In the first step, we identify potential BPOs that could indicate fairness or unfairness. Our analysis is guided by Pohl et al.\cite{pohl2022discrimination} to determine which BPOs can be derived from the MIMIC-IV ED dataset \cite{johnson2023mimic}. Not all BPOs are directly observable in real-world data. Therefore, we assess them based on their presence in the event log, the relevancy of measuring fairness, and whether the outcomes can be meaningfully quantified. Based on \cite{pohl2022discrimination}, seven Business Process Outcomes (BPOs) can be employed to assess fairness. Table \ref{tab:aspect} presents the mapping of the selected process aspects. From this mapping, four outcomes are identifiable here for fairness analysis in the process model: time, re-do, deviation, and decision. Time denotes the duration between the first and the last activity of each case. Re-do refers to an activity that is executed more than once within a single case. Deviation occurs when a process instance deviates from the normal or expected model \cite{pohl2022discrimination}. Finally, decision indicates a point in the process where a choice is made that influences subsequent execution process \cite{Dumas2018Fundamentals}.

\vspace{-1em}
\begin{table}[ht]
\centering
\caption{Business Process Outcomes and Fairness Judgement}
\label{tab:aspect}
\scriptsize 
\renewcommand{\arraystretch}{1.1} 
\resizebox{0.7\textwidth}{!}{ 
\begin{tabular}{|l|c|c|c|c|}
\hline
\textbf{BPOs \cite{pohl2022discrimination}} & \textbf{In Event Log} & \textbf{Relevant} & \textbf{Measurable} & \textbf{Judgement} \\
\hline
\textbf{Time} & Fully present & High & Yes & \cellcolor{gray!30}High \\
\hline
\textbf{Re-do} & Fully present & Medium & Limited & \cellcolor{gray!30}Acceptable \\
\hline
\textbf{Deviation} & Fully present & Medium & Yes & \cellcolor{gray!30}Acceptable \\
\hline
\textbf{Resource allocation} & Not present & Low & No & Insufficient \\
\hline
\textbf{Decision} & Fully present & Medium & Yes & \cellcolor{gray!30}Acceptable \\
\hline
\textbf{Workload} & Not present & Low & No & Insufficient \\
\hline
\textbf{Task complexity} & Not present & Low & Yes & Insufficient \\
\hline
\end{tabular}
}
\end{table}

\vspace{-1em}

The content of Table~\ref{tab:aspect} is derived as follows. For each BPO outcome, we determine its presence: \textit{Fully present} when the information explicitly exists or can be obtained through data processing, and \textit{Not present} if it cannot be derived from the event log. 
This evaluation is based on expert-driven assessment, as no ground truth labels are available for these outcomes. The classification further supports the selection of BPOs that are most feasible for fairness-aware process mining.
\R{1.2}{All of this assessment is based on the availability of each outcome in the empirical data of MIMIC-IV ED \cite{johnson2024mimiciv}.}
The outcome \emph{Time} is fully present as \R{1.8}{an} attribute in most event logs. As it is quantified, fairness can be assessed based on processing time. Therefore, this outcome is judged as High for fairness evaluation. The \emph{Re-do} outcome is also fully present in the event log. However, the fairness implications require deeper exploration, resulting in medium relevance. 
Both \emph{Deviation} and \emph{Decision} are clearly captured in the event log. For relevance, \R{1.9}{these outcomes} require further investigation, such as identifying the extent of deviations and decision distributions. Their measurability is high.
\R{1.2}{Together, these four outcomes capture the core fairness-relevant dimensions of the triage process, namely temporal equity (time), procedural consistency (re-do, deviation), and clinical judgment (decision), all of which are directly observable in emergency care workflows and regulated under frameworks such as EMTALA \cite{zibulewsky2001emergency}.}
In contrast, \emph{Resource allocation} and \emph{Workload} are either not present or only weakly represented in the event log. 
These outcomes, along with \emph{Task complexity}, are therefore considered insufficient for fairness evaluation in this study.


\vspace{-1em}
\subsection{Business Process Outcome Measurement}

\R{1.7}{To quantify each BPO, we first conducted a time-based analysis.} \emph{Time} is the first BPO used to indicate the presence of potential unfairness. In this study, it is measured as the duration between the start and end times of each sequential activity within a single case ID.

Next, we focus on \emph{re-do}. First, we counted the total number of re-do activities occurring within each case ID. We then defined rules to classify which re-do activities represent waste and which are clinically necessary. In healthcare settings, certain activities may occur multiple times due to medical necessity and therefore should not be categorized as waste. Based on this classification scheme (see the extended version \cite{andreswari2026fairnesshealthcareprocessesquantitative}), we summed the number of re-do activities and calculated the percentage of waste relative to the total number of events in each case ID. 

Afterwards, \emph{deviation} is defined as the extent to which process executions deviate from the process model. To assess \emph{deviation}, we applied token-based replay~\cite{van2012replaying} to compare the MIMICEL event log with a process model discovered using the Heuristic Miner with a dependency threshold of 0.8. For each case, we measured fitness based on produced, consumed, remaining, and missing tokens.

The final outcome is \emph{decision}, which was analyzed based on discharge disposition, with patients categorized into five groups.
From the \emph{decision} perspective, we examined how discharge decisions were made across different acuity levels.

\section{Results}
\label{sec:result}

\begin{figure}
    \centering
    \includegraphics[width=0.7\linewidth]{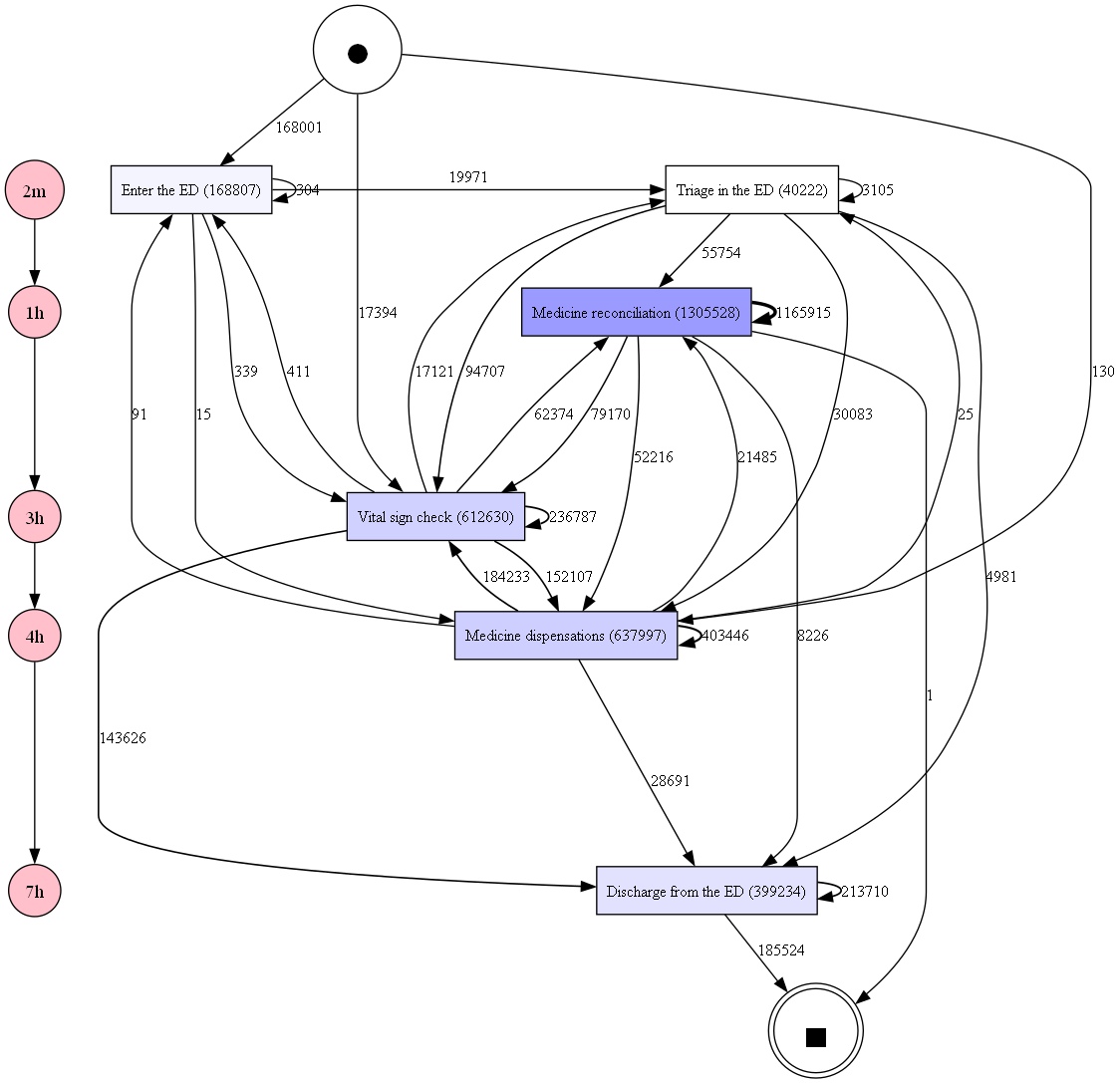}
    \caption{MIMIC-IV ED}
    \label{fig:mimiceltimeline}
\end{figure}

This section presents the results of our fairness analysis for each potential BPOs using the MIMIC-IV ED dataset provided by Johnson et al. \cite{johnson2023mimic} through the PhysioNet platform \cite{goldberger2000physiobank}. 
Figure~\ref{fig:mimiceltimeline} presents an overview of the triage process as mined using the timeline based process discovery~\cite{rubensson2025timeline}. 



\vspace{-1em}

\subsection{Bridging Empirical and Conceptual}

We use the MIMICEL dataset \cite{wei2023mimicel}, which consists of $7{,}422{,}277$ rows with $413{,}893$ cases. The distribution of age groups is: $166{,}965$ in the \textit{until-45} group, $135{,}971$ in the \textit{until-65} group, and $110{,}957$ in the \textit{older} group. The dataset includes $224{,}578$ females and $189{,}315$ males. \R{1.3}{Given the large sample size, statistical significance is expected across most tests and should not be interpreted as evidence of practically meaningful differences. We therefore rely primarily on effect size to assess the magnitude and practical relevance of observed disparities.}

In our analysis, we conduct the Kruskal–Wallis test to determine significant differences of race, gender, age, language, and insurance on \emph{time}, \emph{re-do} and \emph{deviation} for each ESI level, as shown in Tables~\ref{tab:kw_time}–\ref{tab:kw_deviation}. For \emph{decision}, we conduct a chi-square test to quantify the magnitude of differences between groups (see Table~\ref{tab:kw_decision}), as the \emph{decision} variable is categorical and represents discharge outcomes in the emergency department. For these tests, a p-value below the conventional threshold of 0.05 was considered statistically significant, indicating a low probability that the observed differences occurred by chance.

\begin{table}[ht]
\centering
\footnotesize
\renewcommand{\arraystretch}{0.9}
\caption{Kruskal--Wallis test and effect sizes ($\varepsilon^2$) across acuity levels for \emph{time}}
\label{tab:kw_time}
\resizebox{0.6\textwidth}{!}{ 
\begin{tabular}{c l c c c l}
\hline
\textbf{Acuity} & \textbf{Attribute} & \textbf{p-value} & $\boldsymbol{\varepsilon^2}$ & \textbf{Significant} & \textbf{Interpretation} \\
\hline
1 & Race       & $<0.001$ & 0.0004 & Yes & Negligible \\
1 & Age group  & $<0.001$ & 0.0035 & Yes & Negligible \\
1 & Gender     & $<0.001$ & 0.0002 & Yes & Negligible \\
1 & Insurance  & $<0.001$ & 0.0029 & Yes & Negligible \\
1 & Language   & 0.006    & $\sim$0.0000 & Yes & Negligible \\
\hline
2 & Race       & $<0.001$ & 0.0006 & Yes & Negligible \\
2 & Age group  & $<0.001$ & 0.0059 & Yes & Negligible \\
2 & Gender     & $<0.001$ & 0.0002 & Yes & Negligible \\
2 & Insurance  & $<0.001$ & 0.0046 & Yes & Negligible \\
2 & Language   & $<0.001$ & 0.0010 & Yes & Negligible \\
\hline
3 & Race       & $<0.001$ & 0.0014 & Yes & Negligible \\
3 & Age group  & $<0.001$ & 0.0097 & Yes & Negligible \\
3 & Gender     & $<0.001$ & 0.0004 & Yes & Negligible \\
3 & Insurance  & $<0.001$ & 0.0051 & Yes & Negligible \\
3 & Language   & $<0.001$ & 0.0015 & Yes & Negligible \\
\hline
4 & Race       & $<0.001$ & 0.0022 & Yes & Negligible \\
4 & Age group  & $<0.001$ & 0.0175 & Yes & Small \\
4 & Gender     & $<0.001$ & 0.0010 & Yes & Negligible \\
4 & Insurance  & $<0.001$ & 0.0022 & Yes & Negligible \\
4 & Language   & $<0.001$ & 0.0016 & Yes & Negligible \\
\hline
5 & Race       & $<0.001$ & 0.0041 & Yes & Negligible \\
5 & Age group  & $<0.001$ & 0.0285 & Yes & Small \\
5 & Gender     & 0.655    & $\sim$0.0000 & No  & Negligible \\
5 & Insurance  & $<0.001$ & 0.0068 & Yes & Negligible \\
5 & Language   & $<0.001$ & 0.0075 & Yes & Negligible \\
\hline
\end{tabular}
}
\end{table}


In addition, we calculated effect sizes using the epsilon-squared ($\varepsilon^2$) measure for \emph{time}, \emph{re-do} and \emph{deviation}, and Cramér’s V for \emph{decision} outcomes. 
$\varepsilon^2$ values of <0.01, 0.06, 0.14, and >0.14 are interpreted as negligible, small, medium, and large effects, respectively, following commonly adopted conventions for non-parametric effect sizes. Similarly, Cramér’s V values of 0.1, 0.3, 0.5, and >0.5 are considered small, medium, large, and very large effects. These measures provide insights into the magnitude of differences. 

Table~\ref{tab:kw_time} shows that time is statistically significant, with a small effect size, only for acuity levels 4 and 5 for the \emph{age} attribute. This indicates that, in terms of \emph{time}, differentiation across age groups occurs primarily among patients with lower severity levels. This also suggests that patients in different age groups within non-urgent cases are treated differently in terms of duration.

\begin{table}[ht]
\centering
\footnotesize
\renewcommand{\arraystretch}{0.9}
\caption{Kruskal--Wallis test and effect sizes ($\varepsilon^2$) across acuity levels for re-do} 
\label{tab:kw_redo} 
\resizebox{0.6\textwidth}{!}{
\begin{tabular}{c l c c c l}
\hline
\textbf{Acuity} & \textbf{Attribute} & \textbf{p-value} & $\boldsymbol{\varepsilon^2}$ & \textbf{Significant} & \textbf{Interpretation} \\
\hline
1 & Race       & 0.138    & 0.0002 & No  & Negligible \\
1 & Age group  & $<0.001$ & 0.0089 & Yes & Negligible \\
1 & Gender     & $<0.001$ & 0.0007 & Yes & Negligible \\
1 & Insurance  & $<0.001$ & 0.0255 & Yes & Small \\
1 & Language   & $<0.001$ & 0.0228 & Yes & Small \\
\hline
2 & Race       & $<0.001$ & 0.0012 & Yes & Negligible \\
2 & Age group  & $<0.001$ & 0.0059 & Yes & Negligible \\
2 & Gender     & $<0.001$ & 0.0002 & Yes & Negligible \\
2 & Insurance  & $<0.001$ & 0.0439 & Yes & Small \\
2 & Language   & $<0.001$ & 0.0420 & Yes & Small \\
\hline
3 & Race       & $<0.001$ & 0.0015 & Yes & Negligible \\
3 & Age group  & $<0.001$ & 0.0004 & Yes & Negligible \\
3 & Gender     & $<0.001$ & 0.0021 & Yes & Negligible \\
3 & Insurance  & $<0.001$ & 0.0311 & Yes & Small \\
3 & Language   & $<0.001$ & 0.0294 & Yes & Small \\
\hline
4 & Race       & $<0.001$ & 0.0023 & Yes & Negligible \\
4 & Age group  & $<0.001$ & 0.0008 & Yes & Negligible \\
4 & Gender     & $<0.001$ & 0.0032 & Yes & Negligible \\
4 & Insurance  & $<0.001$ & 0.0032 & Yes & Negligible \\
4 & Language   & $<0.001$ & 0.0033 & Yes & Negligible \\
\hline
5 & Race       & 0.798    & $\sim$0.0000 & No  & Negligible \\
5 & Age group  & 0.236    & $\sim$0.0008 & No  & Negligible \\
5 & Gender     & 0.362    & $\sim$0.0000 & No  & Negligible \\
5 & Insurance  & --       & --     & --  & Not tested \\
5 & Language   & 0.220    & $\sim$0.0005 & No  & Negligible \\
\hline
\end{tabular}
}
\end{table}

While the time-based results highlight differences in process duration, it is also essential to investigate whether these differences are reflected in process behavior. We therefore examined whether \emph{re-do} activities were significantly associated with gender, race, age, insurance, and language across ESI levels 1–4 using the Kruskal–Wallis test (Table \ref{tab:kw_redo}). For level 5, none of the attributes were statistically significant. As shown in Table~\ref{tab:kw_redo}, insurance and language show significant differences with small effect sizes at acuity levels 1, 2, and 3. This suggests that insurance and language are systematically associated with differences in process inefficiency across demographic groups and are consistently linked to higher variation in \emph{re-do} behavior among patients with higher severity levels.

\begin{table}[ht] 
\centering 
\footnotesize 
\renewcommand{\arraystretch}{0.9} 
\caption{Kruskal--Wallis test and effect sizes ($\varepsilon^2$) across acuity levels for deviation} 
\label{tab:kw_deviation} 
\resizebox{0.6\textwidth}{!}{
\begin{tabular}{c l c c c l} 
\hline \textbf{Acuity} & \textbf{Attribute} & \textbf{p-value} & $\boldsymbol{\varepsilon^2}$ & \textbf{Significant} & \textbf{Interpretation} \\ 
\hline 
1 & Race & $<0.001$ & 0.0035 & Yes & Negligible \\ 
1 & Age group & $<0.001$ & 0.0443 & Yes & Small \\ 
1 & Gender & 0.381 & $\sim$0.0000 & No & Negligible \\ 
1 & Insurance & $<0.001$ & 0.0924 & Yes & Medium \\ 
1 & Language & $<0.001$ & 0.0701 & Yes & Medium \\ 
\hline 
2 & Race & $<0.001$ & 0.0079 & Yes & Negligible \\ 
2 & Age group & $<0.001$ & 0.0877 & Yes & Medium \\ 
2 & Gender & $<0.001$ & 0.0006 & Yes & Negligible \\ 
2 & Insurance & $<0.001$ & 0.1086 & Yes & Medium \\ 
2 & Language & $<0.001$ & 0.0770 & Yes & Medium \\ 
\hline 
3 & Race & $<0.001$ & 0.0145 & Yes & Small \\ 
3 & Age group & $<0.001$ & 0.1316 & Yes & Medium \\ 
3 & Gender & $<0.001$ & 0.0036 & Yes & Negligible \\ 
3 & Insurance & $<0.001$ & 0.1554 & Yes & Large \\ 
3 & Language & $<0.001$ & 0.1375 & Yes & Medium \\ 
\hline 
4 & Race & $<0.001$ & 0.0088 & Yes & Negligible \\ 
4 & Age group & $<0.001$ & 0.1243 & Yes & Medium \\ 
4 & Gender & $<0.001$ & 0.0056 & Yes & Negligible \\ 
4 & Insurance & $<0.001$ & 0.0432 & Yes & Small \\ 
4 & Language & $<0.001$ & 0.0399 & Yes & Small \\ 
\hline 
5 & Race & 0.015 & 0.0079 & Yes & Negligible \\ 
5 & Age group & $<0.001$ & 0.0694 & Yes & Medium \\ 
5 & Gender & 0.657 & $\sim$0.0000 & No & Negligible \\ 
5 & Insurance & -- & -- & -- & Not tested \\ 
5 & Language & $<0.001$ & 0.0173 & Yes & Small \\ 
\hline 
\end{tabular} 
}
\end{table}

Moreover, for the \emph{deviation} outcome, statistically significant differences with higher effect sizes were observed for acuity level 3 across all attributes (p < 0.05), with a large effect size for insurance. In contrast, at acuity levels 1, 2, 4, and 5, the effect sizes were generally lower (Table \ref{tab:kw_deviation}). Across all acuity levels, race and gender consistently indicate the lowest effect sizes, followed by age, while insurance and language tend to show similar and relatively higher effect sizes. For acuity level 5, the insurance attribute was not tested due to insufficient group size (n < 30). Overall, deviation is more strongly associated with differences in age, insurance, and language than with race or gender. 


\begin{table}[ht]
\centering
\footnotesize
\renewcommand{\arraystretch}{0.9}
\caption{Chi-square test and effect sizes (Cramér’s V) across acuity levels for discharge decision}
\label{tab:kw_decision}
\resizebox{0.6\textwidth}{!}{
\begin{tabular}{c l c c c l}
\hline
\textbf{Acuity} & \textbf{Attribute} & \textbf{p-value} & \textbf{Cramér’s V} & \textbf{Significant} & \textbf{Interpretation} \\
\hline
1 & Race       & $<0.001$ & 0.049  & Yes & Small \\
1 & Age group  & $<0.001$ & 0.29   & Yes & Medium \\
1 & Gender     & $<0.001$ & 0.0702 & Yes & Small \\
1 & Insurance  & $<0.001$ & 0.37   & Yes & Large \\
1 & Language   & $<0.001$ & 0.49   & Yes & Large \\
\hline
2 & Race       & $<0.001$ & 0.06   & Yes & Small \\
2 & Age group  & $<0.001$ & 0.26   & Yes & Medium \\
2 & Gender     & $<0.001$ & 0.0549 & Yes & Small \\
2 & Insurance  & $<0.001$ & 0.33   & Yes & Large \\
2 & Language   & $<0.001$ & 0.42   & Yes & Large \\
\hline
3 & Race       & $<0.001$ & 0.07   & Yes & Small \\
3 & Age group  & $<0.001$ & 0.21   & Yes & Medium \\
3 & Gender     & $<0.001$ & 0.0433 & Yes & Small \\
3 & Insurance  & $<0.001$ & 0.37   & Yes & Large \\
3 & Language   & $<0.001$ & 0.49   & Yes & Large \\
\hline
4 & Race       & 0.179    & 0.0164 & No  & Small \\
4 & Age group  & $<0.001$ & 0.0875 & Yes & Small \\
4 & Gender     & 0.413    & 0.0119 & No  & Small \\
4 & Insurance  & $<0.001$ & 0.29   & Yes & Medium \\
4 & Language   & $<0.001$ & 0.37   & Yes & Large \\
\hline
5 & Race       & 0.808    & 0.046  & No  & Small \\
5 & Age group  & $<0.001$ & 0.1057 & Yes & Medium \\
5 & Gender     & 0.122    & 0.0628 & No  & Small \\
5 & Insurance  & $<0.001$ & 0.5024 & Yes & Very large \\
5 & Language   & $<0.001$ & 0.394  & Yes & Large \\
\hline
\end{tabular}
}
\end{table}


The chi-square test for the \emph{decision} outcome reveals similar patterns for acuity levels 1–3, where all attributes are statistically significant (Table \ref{tab:kw_decision}). Also race and gender show lower effect sizes than age, while insurance and language indicate the largest effect sizes. For acuity levels 4 and 5, race and gender are not statistically significant, whereas insurance and language remain associated with the highest effect sizes. Notably, for acuity level 5, insurance demonstrates a very large effect size compared to other demographic attributes. Overall, these findings suggest that discharge decisions in the emergency department are more strongly associated with insurance and language across all acuity levels, while race and gender show limited association with disparities in decision outcomes.

\subsection{Characterizing by Justice Dimension}
\vspace{-0.5em}

This section integrates the results of the BPOs into relevant justice dimensions. Based on the analysis, we characterize each BPO by aligning it with a corresponding justice principle \cite{gilliland1993perceived} and providing the rationale, as summarized in Table~\ref{tab:justice_summary}. After evaluating the statistical significance of the BPOs, we identified \emph{time}, \emph{re-do}, \emph{deviation}, and \emph{decision} as potential indicators of differences in the process. Specifically, \emph{decision} is associated with distributive justice, as it reflects whether patients receive appropriate discharge decisions based on their medical needs, \R{1.1, 1.4}{operationalized through Cramér's V across demographic groups.} In contrast, \emph{time}, \emph{deviation}, and \emph{re-do} relate to process consistency and adherence to procedures, which align with procedural justice, \R{1.1, 1.4}{operationalized through epsilon-squared ($\varepsilon^2$).} Lastly, \emph{re-do}, as an outcome related to interaction during care delivery, is also associated with interactional justice, \R{1.1, 1.4}{particularly when repetitions stem from communication barriers as reflected in the language attribute. Where a BPO maps to multiple dimensions, this reflects different aspects of the same process behavior rather than a conflict.}
As highlighted in the results, acuity levels 1–3 show significant differences across age, insurance, and language, reinforcing the importance of fairness considerations. In contrast, for the \emph{time} outcome, statistically significant differences are observed only at acuity levels 4–5 for the age attribute. 
These findings support the conceptual linkage between BPOs and justice dimensions, providing insights for further fairness-aware process mining research. Integrating empirical findings into justice dimensions not only offers a structured framework for fairness analysis but also provides practical insights for improving emergency healthcare processes. By identifying how BPOs reflect distributive, procedural, and interactional justice, stakeholders can better monitor, audit, and redesign triage workflows to reduce disparities.
\begin{table}[ht]
\centering
\footnotesize
\caption{Summary of results by organizational justice dimension}
\label{tab:justice_summary}
\renewcommand{\arraystretch}{1.05}
\resizebox{0.8\textwidth}{!}{
\begin{tabular}{p{2cm} p{3.2cm} p{2cm} p{2cm} p{2.2cm}}
\hline
\textbf{Justice type} & \textbf{BPO} & \textbf{Acuity level(s)} & \textbf{Key attributes} & \textbf{Effect size} \\
\hline
Distributive & Decision & 1--5 &
\shortstack[l]{Age\\Insurance\\Language} &
\shortstack[l]{Medium--\\very large} \\
\hline
 & Time & 4--5 &
\shortstack[l]{Age} &
Negligible--small \\
\cline{2-5}
Procedural & Deviation & \shortstack[l]{1--3 (strong), \\4--5 (weak)} &
\shortstack[l]{Age\\Insurance\\Language} &
Small--large \\
\cline{2-5}
 & Re-do & 1--3 &
\shortstack[l]{Insurance\\Language} &
\shortstack[l]{Negligible--\\small} \\
\hline
Interactional & Re-do (interaction-related) & 1--3 &
\shortstack[l]{Language} &
Negligible--small \\
\hline
\end{tabular}
}
\end{table}


\subsection{Discussion}



The findings suggest that fairness disparities might exist in the triage process, particularly among patients assigned the same ESI level. Our results show that patients at acuity levels 1–3 experience significant differences across \emph{re-do}, \emph{deviation}, and \emph{decision}, particularly with respect to age, insurance, and language. Although this case study focuses on triage and the Medical Screening Examination (MSE), which is regulated under EMTALA \cite{zibulewsky2001emergency} to ensure equal access to emergency care, disparities still emerge, with patient insurance status remaining a major influencing factor. In addition, patients with acuity level 4 and 5 also show disparities associated with age factor, particularly in terms of \emph{time} outcome. 
This pattern suggests that age-related differences in waiting time may reflect prioritization practices not only based on acuity but also influenced by perceived vulnerability.


While Lauridsen et al.~\cite{lauridsen2020emergency} identified sub-acute categories as particularly vulnerable in triage decisions, our results show that significant differences are more evident for acuity levels 1–3, especially with respect to age, insurance, and language. This apparent difference reflects variations in triage systems rather than a contradiction in findings. In both cases, disparities emerge in situations characterized by high clinical process complexity, where patient pathways may diverge. In our setting, such conditions occur not only among lower-acuity and sub-acute patients but also extend to high-acuity cases, where coordination and communication demands are substantial. These findings further demonstrate that disparities span a broader range of clinical severities when examined from a process-oriented perspective.



On the one hand, our observations raise concerns about distributive, procedural and interactive justice in triage practices, even in critical emergency care. On the other hand, given the absence of comorbidity or prior condition data in MIMIC-IV ED, these findings warrant cautious interpretation. As fairness cannot be directly observed, our statistical analysis is an analysis of differences \R{1.1}{that serves as a diagnostic foundation for developing targeted fairness interventions in triage workflows.} Mind that various other factors not available in our dataset could be contributing to these differences, relating to medical conditions, and operational workload.

Although this study relies on a single dataset from one healthcare institution, the analyzed triage process is aligned with the ESI method, which is widely applied in emergency departments worldwide. The medical center contributing to the MIMIC-IV dataset also operates under formal accreditation standards that support nationwide ranking achievements. Thus, despite its context-specific scope, the process design reflects internationally recognized triage principles.

Future work should explore integrating additional patient attributes such as medical records, and socioeconomic status to more fully contextualize outcomes, as well as the time of emergency visits and the workload of medical staff. Additionally, linking event logs with other clinical systems that provide more detailed patient records may help uncover the root causes of process-based disparities. \R{3.4}{Furthermore, applying additional fairness metrics such as demographic parity, equalized odds, and disparate impact \cite{pessach2022review}, \cite{garg2020fairness} could complement the distributional analysis presented here and provide a more comprehensive assessment of fairness in triage processes.} Developing fairness-aware process monitoring tools for emergency healthcare could support real-time auditing and assist decision-makers in providing more equitable treatment paths.

\section{Conclusion}
\label{sec:conclusion}
\vspace{-1em}
This study demonstrates the importance of integrating fairness considerations into process mining practices, particularly in critical healthcare settings such as emergency triage. By leveraging the MIMIC-IV ED dataset and grounding our analysis in justice theory, we identified statistically significant differences in business process outcomes related to age, race, gender, insurance and language especially among high-acuity to sub-acute patient categories (level 1-3). 
This raises concerns about procedural, distributive and interactional justice in healthcare workflows. By mapping empirical disparities to justice dimensions, our work provides a structured approach for evaluating fairness in real-life event logs, supporting the broader agenda of responsible and ethics-aware process mining.
\R{1.1, 2.1}{Although the operationalization of BPOs is instantiated on the MIMIC-IV ED dataset \cite{johnson2024mimiciv}, the proposed framework is transferable to other triage settings that adopt the ESI method or similar acuity-based classification systems, as these share the same fundamental process structure. From an information systems perspective, this work contributes a diagnostic framework for fairness in process-aware information systems \cite{van2009process}, supporting the design of more equitable healthcare information systems and advancing the agenda of responsible process mining \cite{Mannhardt2022Responsible}.}

\begin{credits}
\subsubsection{\ackname} This research was supported by the Einstein Foundation Berlin under grant EPP-2019-524, by the Federal Ministry of Research, Technology and Space under the grant 16DII133, and by Deutsche Forschungsgemeinschaft under grants 496119880 (VisualMine), 531115272 (ProImpact), SFB 1404/2 (FONDA), and Elsa Neumann Stipendium (H78027).
\vspace{-1em}
\end{credits}

%
%
%
\bibliographystyle{splncs04}
\bibliography{bibliography}

@article{rojas2016process,
  title={Process mining in healthcare: A literature review},
  author={Rojas, Eric and Munoz-Gama, Jorge and Sep{\'u}lveda, Marcos and Capurro, Daniel},
  journal={Journal of biomedical informatics},
  volume={61},
  pages={224--236},
  year={2016},
  publisher={Elsevier}
}

@inproceedings{peeperkorn2025achieving,
  title={Achieving Group Fairness Through Independence in Predictive Process Monitoring},
  author={Peeperkorn, Jari and De Vos, Simon},
  booktitle={International Conference on Advanced Information Systems Engineering},
  pages={185--203},
  year={2025},
  organization={Springer}
}

@inproceedings{muskan2024extending,
  title={Extending genetic process discovery to reveal unfairness in processes},
  author={Muskan and Mannhardt, Felix and van Dongen, Boudewijn},
  booktitle={International Conference on Process Mining},
  pages={751--763},
  year={2024},
  organization={Springer}
}

@inproceedings{de2024achieving,
  title={Achieving Fairness in Predictive Process Analytics via Adversarial Learning},
  author={de Leoni, Massimiliano and Padella, Alessandro},
  booktitle={International Conference on Cooperative Information Systems},
  pages={346--354},
  year={2024},
  organization={Springer}
}

@article{joshi1989measurement,
  title={The measurement of fairness or equity perceptions of management information systems users},
  author={Joshi, Kailash},
  journal={MIS quarterly},
  volume={13},
  number={3},
  pages={343--358},
  year={1989},
  publisher={Management Information Systems Research Center, University of Minnesota}
}

@incollection{van2009process,
  title={Process-aware information systems: Lessons to be learned from process mining},
  author={Van der Aalst, Wil MP},
  booktitle={Transactions on Petri Nets and Other Models of Concurrency II: Special Issue on Concurrency in Process-Aware Information Systems},
  pages={1--26},
  year={2009},
  publisher={Springer}
}

@incollection{dorleon2023fapfid,
  title={FAPFID: a fairness-aware approach for protected features and imbalanced data},
  author={Dorleon, Ginel and Megdiche, Imen and Bricon-Souf, Nathalie and Teste, Olivier},
  booktitle={Transactions on Large-Scale Data-and Knowledge-Centered Systems LIII},
  pages={107--125},
  year={2023},
  publisher={Springer}
}

@inproceedings{garg2020fairness,
  title={Fairness metrics: A comparative analysis},
  author={Garg, Pratyush and Villasenor, John and Foggo, Virginia},
  booktitle={2020 IEEE international conference on big data (Big Data)},
  pages={3662--3666},
  year={2020},
  organization={IEEE}
}

@article{fernandes2020predicting,
  title={Predicting Intensive Care Unit admission among patients presenting to the emergency department using machine learning and natural language processing},
  author={Fernandes, Marta and Mendes, Ruben and Vieira, Susana M and Leite, Francisca and Palos, Carlos and Johnson, Alistair and Finkelstein, Stan and Horng, Steven and Celi, Leo Anthony},
  journal={PloS one},
  volume={15},
  number={3},
  pages={e0229331},
  year={2020},
  publisher={Public Library of Science San Francisco, CA USA}
}

@misc{andreswari2026fairnesshealthcareprocessesquantitative,
      title={Fairness in Healthcare Processes: A Quantitative Analysis of Decision Making in Triage}, 
      author={Rachmadita Andreswari and Stephan A. Fahrenkrog-Petersen and Jan Mendling},
      year={2026},
      eprint={2601.11065},
      archivePrefix={arXiv},
      primaryClass={cs.CY},
      
}

@inproceedings{zibulewsky2001emergency,
  title={The Emergency Medical Treatment and Active Labor Act (EMTALA): what it is and what it means for physicians},
  author={Zibulewsky, Joseph},
  booktitle={Baylor University Medical Center Proceedings},
  volume={14},
  number={4},
  pages={339--346},
  year={2001},
  organization={Taylor \& Francis}
}

@article{gilboy2012emergency,
  title={Emergency Severity Index (ESI): a triage tool for emergency department care, version 4},
  author={Gilboy, Nicki and Tanabe, Paula and Travers, Debbie and Rosenau, Alexander M and others},
  journal={Implementation handbook},
  volume={2012},
  pages={12--0014},
  year={2012}
}

@article{cappelen2006responsibility,
  title={Responsibility, fairness and rationing in health care},
  author={Cappelen, Alexander W and Norheim, Ole Frithjof},
  journal={Health policy},
  volume={76},
  number={3},
  pages={312--319},
  year={2006},
  publisher={Elsevier}
}

@inproceedings{chakraborty2024setting,
  title={Setting up a just and fair icu triage process during a pandemic: a systematic review},
  author={Chakraborty, Rhyddhi and Achour, Nebil},
  booktitle={Healthcare},
  volume={12},
  number={2},
  pages={146},
  year={2024},
  organization={MDPI}
}

@article{kern2024navigating,
  title={Navigating the moral maze: a literature review of ethical values in business process management},
  author={Kern, Christopher Julian and Poss, Leo and Kroenung, Julia and Sch{\"o}nig, Stefan},
  journal={Business Process Management Journal},
  volume={30},
  number={8},
  pages={343--370},
  year={2024},
  publisher={Emerald Publishing Limited}
}

@article{varkey2021principles,
  title={Principles of clinical ethics and their application to practice},
  author={Varkey, Basil},
  journal={Medical Principles and Practice},
  volume={30},
  number={1},
  pages={17--28},
  year={2021},
  publisher={S. Karger AG Basel, Switzerland}
}

@inproceedings{qafari2019fairness,
  title={Fairness-aware process mining},
  author={Qafari, Mahnaz Sadat and Van der Aalst, Wil},
  booktitle={On the Move to Meaningful Internet Systems: OTM 2019 Conferences: Confederated International Conferences: CoopIS, ODBASE, C\&TC 2019, Rhodes, Greece, October 21--25, 2019, Proceedings},
  pages={182--192},
  year={2019},
  organization={Springer}
}

@inproceedings{pohl2022discrimination,
  title={Discrimination-aware process mining: a discussion},
  author={Pohl, Timo and Qafari, Mahnaz Sadat and van der Aalst, Wil MP},
  booktitle={ICPM Workshops},
  pages={101--113},
  year={2022},
  organization={Springer}
}

@article{ahmed2024ethical,
  title={Ethical triage in public health emergency facilities: distributive justice--a decision model},
  author={Ahmed, Shamsuddin and Alsisi, Rayan Hamza},
  journal={Kybernetes},
  year={2024},
  publisher={Emerald Publishing Limited}
}

@article{holzer2025role,
  title={The role of social justice in triage revisited: a threshold conception},
  author={Holzer, Felicitas and Biller-Andorno, Nikola and Baumann, Holger},
  journal={Medicine, Health Care and Philosophy},
  volume={28},
  number={1},
  pages={161--169},
  year={2025},
  publisher={Springer}
}

@article{ghanbari2021fair,
  title={Fair prioritization of casualties in disaster triage: a qualitative study},
  author={Ghanbari, Vahid and Ardalan, Ali and Zareiyan, Armin and Nejati, Amir and Hanfling, Dan and Bagheri, Alireza and Rostamnia, Leili},
  journal={BMC emergency medicine},
  volume={21},
  pages={1--9},
  year={2021},
  publisher={Springer}
}

@inproceedings{broome1990fairness,
  title={Fairness},
  author={Broome, John},
  booktitle={Proceedings of the Aristotelian society},
  volume={91},
  pages={87--101},
  year={1990},
  organization={JSTOR}
}

@article{lauridsen2020emergency,
  title={Emergency care, triage, and fairness},
  author={Lauridsen, Sigurd},
  journal={Bioethics},
  volume={34},
  number={5},
  pages={450--458},
  year={2020},
  publisher={Wiley Online Library}
}

@misc{wei2023mimicel,
  author       = {Wei, Jiayuan and He, Zhen and Ouyang, Chen and Moreira, Claudio},
  title        = {{MIMICEL: MIMIC-IV Event Log for Emergency Department (version 2.1.0)}},
  year         = {2023},
  note         = {PhysioNet. RRID:SCR\_007345},
  howpublished = {%\url{https://doi.org/10.13026/c9yj-1t90}}
}

@misc{johnson2024mimiciv,
  author       = {Johnson, Alistair and Bulgarelli, Luca and Pollard, Tom and Gow, Benjamin and Moody, Benjamin and Horng, Steven and Celi, Leo Anthony and Mark, Roger},
  title        = {{MIMIC-IV (version 3.1)}},
  year         = {2024},
  note         = {PhysioNet. RRID:SCR\_007345},
  howpublished = {%\url{https://doi.org/10.13026/kpb9-mt58}}
}

@misc{johnson2023mimic,
  author       = {Johnson, Alistair and Bulgarelli, Luca and Pollard, Tom and Celi, Leo Anthony and Mark, Roger and Horng, Steven},
  title        = {{MIMIC-IV-ED (version 2.2)}},
  year         = {2023},
  note         = {PhysioNet. RRID:SCR\_007345},
  howpublished = {%\url{https://doi.org/10.13026/5ntk-km72}}
}

@article{chou2007gender,
  title={Gender and racial disparities in the management of diabetes mellitus among Medicare patients},
  author={Chou, Ann F and Brown, Arleen F and Jensen, Roxanne E and Shih, Sarah and Pawlson, Greg and Scholle, Sarah Hudson},
  journal={Women's Health Issues},
  volume={17},
  number={3},
  pages={150--161},
  year={2007},
  publisher={Elsevier}
}

@article{van2012replaying,
  title={Replaying history on process models for conformance checking and performance analysis},
  author={Van der Aalst, Wil and Adriansyah, Arya and Van Dongen, Boudewijn},
  journal={Wiley Interdisciplinary Reviews: Data Mining and Knowledge Discovery},
  volume={2},
  number={2},
  pages={182--192},
  year={2012},
  publisher={Wiley Online Library}
}

@article{buja2014need,
  title={Need and disparities in primary care management of patients with diabetes},
  author={Buja, Alessandra and others},
  journal={BMC Endocrine Disorders},
  volume={14},
  pages={1--8},
  year={2014},
  publisher={Springer}
}

@article{pessach2022review,
  title={A review on fairness in machine learning},
  author={Pessach, Dana and Shmueli, Erez},
  journal={ACM CSUR},
  volume={55},
  number={3},
  pages={1--44},
  year={2022},
  publisher={ACM New York, NY}
}

@article{gilliland1993perceived,
  title={The perceived fairness of selection systems: An organizational justice perspective},
  author={Gilliland, Stephen W},
  journal={Academy of management review},
  volume={18},
  number={4},
  pages={694--734},
  year={1993},
  publisher={Academy of Management Briarcliff Manor, NY 10510}
}

@article{goldberger2000physiobank,
  author       = {Goldberger, Ary L. and Amaral, Luis A. N. and Glass, Leon and Hausdorff, Jeffrey M. and Ivanov, Plamen Ch. and Mark, Roger G. and Stanley, H. Eugene},
  title        = {PhysioBank, PhysioToolkit, and PhysioNet: Components of a new research resource for complex physiologic signals},
  journal      = {Circulation},
  volume       = {101},
  number       = {23},
  pages        = {e215--e220},
  year         = {2000},
  publisher    = {American Heart Association},
  note         = {[Online]},
}

@article{Dumas2018Fundamentals,
   author = {Marlon Dumas and Marcello La Rosa and Jan Mendling and Hajo A. Reijers},
   isbn = {9783662565094},
   month = {3},
   pages = {1-527},
   publisher = {Springer Berlin Heidelberg},
   title = {Fundamentals of business process management: Second Edition},
   year = {2018},
}

@article{Mannhardt2022Responsible,
   author = {Felix Mannhardt},
   issn = {18651356},
   journal = {Lecture Notes in Business Information Processing},
   pages = {373-401},
   publisher = {Springer Science and Business Media Deutschland GmbH},
   title = {Responsible Process Mining},
   volume = {448},
   year = {2022},
}

@article{Aalst2017Responsible,
   author = {Wil M.P. van der Aalst and Martin Bichler and Armin Heinzl},
   issn = {18670202},
   issue = {5},
   journal = {Business \& Information Systems Engineering},
   month = {10},
   pages = {311-313},
   publisher = {Gabler Verlag},
   title = {Responsible Data Science},
   volume = {59},
   year = {2017},
}

@article{rubensson2025timeline,
  title={Timeline-based process discovery},
  author={Rubensson, Christoffer and Kaur, Harleen and Kampik, Timotheus and Mendling, Jan},
  journal={Information Systems},
  pages={102568},
  year={2025},
  publisher={Elsevier}
}

@article{runeson2009guidelines,
  title={Guidelines for conducting and reporting case study research in software engineering},
  author={Runeson, Per and H{\"o}st, Martin},
  journal={Empirical software engineering},
  volume={14},
  number={2},
  pages={131--164},
  year={2009},
  publisher={Springer}
}

@incollection{sapra2023vital,
  title={Vital sign assessment},
  author={Sapra, Amit and Malik, Ahmad and Bhandari, Priyanka},
  booktitle={StatPearls [internet]},
  year={2023},
  publisher={StatPearls Publishing}
}

@article{mashoufi2023data,
  title={Data quality assessment in emergency medical services: an objective approach},
  author={Mashoufi, Mehrnaz and Ayatollahi, Haleh and Khorasani-Zavareh, Davoud and Talebi Azad Boni, Tahere},
  journal={BMC Emergency Medicine},
  volume={23},
  number={1},
  pages={10},
  year={2023},
  issn={1471-227X},
  publisher={BioMed Central}
}

@article{soola2022evaluation,
  title={Evaluation of the factors affecting triage decision-making among emergency department nurses and emergency medical technicians in Iran: a study based on Benner's theory},
  author={Soola, Aghil Habibi and Mehri, Saeid and Azizpour, Islam},
  journal={BMC Emergency Medicine},
  volume={22},
  number={1},
  pages={174},
  year={2022},
  doi={10.1186/s12873-022-00729-y},
  pmid={36303127},
  pmcid={PMC9613063},
  issn={1471-227X},
  publisher={BioMed Central}
}

@article{iserson2007triage,
  title={Triage in medicine, part I: concept, history, and types},
  author={Iserson, Kenneth V and Moskop, John C},
  journal={Annals of emergency medicine},
  volume={49},
  number={3},
  pages={275--281},
  year={2007},
  publisher={Elsevier}
}

\clearpage
\appendix
\section{APPENDIX}
\label{sec:appendix}
\begin{table}[htbp]
\centering
\caption{Race Category Mapping}
\label{tab:race_mapping}
\resizebox{\textwidth}{!}{%
\begin{tabular}{lrl l}
\hline
\textbf{Original Category} & \textbf{Count} & \textbf{New Category} & \textbf{Race Group} \\
\hline
WHITE & 228,077 & WHITE & Caucasian \\
WHITE -- OTHER EUROPEAN & 8,991 & WHITE & Caucasian \\
WHITE -- RUSSIAN & 6,091 & WHITE & Caucasian \\
WHITE -- BRAZILIAN & 1,484 & WHITE & Caucasian \\
PORTUGUESE & 1,457 & WHITE & Caucasian \\
WHITE -- EASTERN EUROPEAN & 1,312 & WHITE & Caucasian \\
HISPANIC OR LATINO & 3,140 & UNKNOWN\_RACE & Delete \\
PATIENT DECLINED TO ANSWER & 624 & UNKNOWN\_RACE & Delete \\
MULTIPLE RACE/ETHNICITY & 281 & UNKNOWN\_RACE & Delete \\
UNABLE TO OBTAIN & 244 & UNKNOWN\_RACE & Delete \\
HISPANIC/LATINO -- PUERTO RICAN & 14,032 & MULTI\_ETHNIC & Multiethnic \\
HISPANIC/LATINO -- DOMINICAN & 8,328 & MULTI\_ETHNIC & Multiethnic \\
UNKNOWN & 7,051 & MULTI\_ETHNIC & Multiethnic \\
HISPANIC/LATINO -- GUATEMALAN & 2,352 & MULTI\_ETHNIC & Multiethnic \\
HISPANIC/LATINO -- SALVADORAN & 1,497 & MULTI\_ETHNIC & Multiethnic \\
HISPANIC/LATINO -- COLUMBIAN & 1,306 & MULTI\_ETHNIC & Multiethnic \\
HISPANIC/LATINO -- MEXICAN & 1,260 & MULTI\_ETHNIC & Multiethnic \\
SOUTH AMERICAN & 1,063 & MULTI\_ETHNIC & Multiethnic \\
HISPANIC/LATINO -- HONDURAN & 1,005 & MULTI\_ETHNIC & Multiethnic \\
HISPANIC/LATINO -- CUBAN & 790 & MULTI\_ETHNIC & Multiethnic \\
HISPANIC/LATINO -- CENTRAL AMERICAN & 788 & MULTI\_ETHNIC & Multiethnic \\
BLACK/AFRICAN AMERICAN & 76,777 & BLACK & Non-Caucasian \\
BLACK/CAPE VERDEAN & 7,635 & BLACK & Non-Caucasian \\
BLACK/AFRICAN & 4,887 & BLACK & Non-Caucasian \\
BLACK/CARIBBEAN ISLAND & 3,674 & BLACK & Non-Caucasian \\
ASIAN -- CHINESE & 7,348 & ASIAN & Non-Caucasian \\
ASIAN & 7,294 & ASIAN & Non-Caucasian \\
ASIAN -- ASIAN INDIAN & 1,565 & ASIAN & Non-Caucasian \\
ASIAN -- SOUTH EAST ASIAN & 1,533 & ASIAN & Non-Caucasian \\
AMERICAN INDIAN/ALASKA NATIVE & 1,037 & NATIVE\_AMERICAN & Non-Caucasian \\
ASIAN -- KOREAN & 786 & ASIAN & Non-Caucasian \\
NATIVE HAWAIIAN OR OTHER PACIFIC ISLANDER & 507 & PACIFIC\_ISLANDER & Non-Caucasian \\
OTHER & 20,736 & OTHER & Other \\
\hline
\end{tabular}
}
\end{table}

\begin{table}[htbp]
\centering
\caption{Language Category Mapping}
\label{tab:language_mapping}
\begin{tabular}{lrl}
\hline
\textbf{Original Category} & \textbf{Count} & \textbf{Language Group} \\
\hline
ENGLISH & 181,615 & ENGLISH \\
SPANISH & 7,364 & NON\_ENGLISH \\
RUSSIAN & 3,488 & NON\_ENGLISH \\
CHINESE & 2,866 & NON\_ENGLISH \\
KABUVERDIANU & 1,994 & NON\_ENGLISH \\
HAITIAN & 1,069 & NON\_ENGLISH \\
PORTUGUESE & 978 & NON\_ENGLISH \\
OTHER & 526 & NON\_ENGLISH \\
VIETNAMESE & 440 & NON\_ENGLISH \\
MODERN GREEK (1453--) & 407 & NON\_ENGLISH \\
ITALIAN & 399 & NON\_ENGLISH \\
ARABIC & 264 & NON\_ENGLISH \\
AMERICAN SIGN LANGUAGE & 237 & NON\_ENGLISH \\
POLISH & 199 & NON\_ENGLISH \\
PERSIAN & 196 & NON\_ENGLISH \\
KOREAN & 133 & NON\_ENGLISH \\
THAI & 108 & NON\_ENGLISH \\
FRENCH & 99 & NON\_ENGLISH \\
AMHARIC & 98 & NON\_ENGLISH \\
UNKNOWN & 222,078 & UNKNOWN \\
\hline
\end{tabular}
\end{table}

\begin{table}[htbp]
\centering
\caption{Insurance Category Mapping}
\label{tab:insurance_mapping}
\begin{tabular}{lrl}
\hline
\textbf{Original Category} & \textbf{Count} & \textbf{Insurance Group} \\
\hline
MEDICARE & 95,009 & Public \\
MEDICAID & 42,437 & Public \\
PRIVATE & 56,357 & Private \\
OTHER & 5,348 & Private \\
UNKNOWN & 225,797 & Unknown \\
NO CHARGE & 4 & Unknown \\
\hline
\end{tabular}
\end{table}

\begin{table}[ht]
\centering
\small
\caption{Mapping of discharge dispositions to discharge groups}
\label{tab:discharge_group}
\begin{tabular}{l r l}
\hline
\textbf{Discharge disposition} & \textbf{Count} & \textbf{Discharge group} \\
\hline
UNKNOWN                        & 289{,}265 & UNKNOWN\\ 
HOME                           & 55{,}540  & HOME\\ 
HOME HEALTH CARE               & 31{,}263  & HOME\\ 
SKILLED NURSING FACILITY       & 20{,}599  & FACILITY \\
REHAB                          & 4{,}405   & FACILITY \\
DIED                           & 3{,}814   & DEATH \\
CHRONIC/LONG TERM ACUTE CARE   & 2{,}710   & FACILITY \\
HOSPICE                        & 2{,}124   & DEATH \\
AGAINST ADVICE                & 1{,}520   & AGAINST\_ADVICE \\
PSYCH FACILITY                & 1{,}089   & FACILITY \\
OTHER FACILITY                & 650       & FACILITY \\
ACUTE HOSPITAL                & 609       & FACILITY \\
ASSISTED LIVING               & 281       & FACILITY \\
HEALTHCARE FACILITY           & 24        & FACILITY \\
\hline
\end{tabular}
\end{table}

\begin{table}[ht]
\centering
\small
\caption{Re-do (rework) classification rules per activity}
\label{tab:rework_rules}
\renewcommand{\arraystretch}{1.15}
\begin{tabular}{p{4.0cm} p{4.0cm} p{4.6cm}}
\hline
\textbf{Activity} & \textbf{Expected frequency per stay ID} & \textbf{Rework classification} \\
\hline
Enter the ED & 1 & If more than one occurrence; potential waste (waste redo) \\
\hline
Triage in the ED & 1 (rare update) &  More than one occurrence indicates potential waste; further assessment required to distinguish clinical redo from waste redo. If acuity changes or increases and the time gap exceeds 30 minutes, the repetition is considered clinically justified (clinical redo), otherwise waste redo \\
\hline
Vital sign check & Multiple & Clinically needed (clinical redo) \\
\hline
Medicine dispensations & Multiple & Clinically needed (clinical redo) \\
\hline
Medicine reconciliation & Multiple (conditional) & Clinically needed (clinical redo)  \\
\hline
Discharge from the ED & 1 & If more than one occurrence; potential waste (waste redo) \\
\hline
\end{tabular}
\end{table}

\end{document}